\documentclass[aps,prl,twocolumn,superscriptaddress]{revtex4-2}
\usepackage{amsmath,amssymb}
\usepackage{graphicx}
\usepackage{times}
\usepackage{amsthm}
\usepackage{amsfonts}
\usepackage[english]{babel}
\usepackage{mathtools}
\usepackage{float}
\usepackage[dvipsnames]{xcolor}
\usepackage{siunitx}

\begin{document}

\title{Spectral fringes without subcycles in Schwinger pair production and Dirac materials}

\author{I.~A.~Aleksandrov}
\affiliation{Department of Physics, Saint Petersburg State University, Universitetskaya Naberezhnaya 7/9, Saint Petersburg 199034, Russia}
\author{M.~A.~Dorodnyi}
\affiliation{Department of Physics, Saint Petersburg State University, Universitetskaya Naberezhnaya 7/9, Saint Petersburg 199034, Russia}
\author{E.~D.~Akimkina}
\affiliation{Department of Physics, Saint Petersburg State University, Universitetskaya Naberezhnaya 7/9, Saint Petersburg 199034, Russia}

\begin{abstract}
Spectral fringes in Schwinger pair creation are usually attributed to structured driving, such as carrier oscillations, pulse trains, or multiple creation events. We show that pronounced fringes can arise even for smooth, carrier-free single-lobe electric-field pulses. Two bell-shaped profiles that are nearly indistinguishable in real time---a Gaussian pulse and a weakly deformed variant---produce qualitatively different longitudinal momentum spectra in the nonadiabatic crossover: the Gaussian spectrum remains smooth, whereas the deformed pulse develops strong fringes as the Keldysh parameter approaches unity. Exact numerical solutions in scalar and spinor QED agree with a semiclassical turning-point analysis and trace the effect to a turning-point dominance transition, where the leading saddle becomes irrelevant and subleading contributions interfere. We demonstrate the same mechanism in a solid-state Schwinger analog described by a gapped two-dimensional Dirac model relevant to epitaxial graphene on SiC, and discuss an energy-resolved pump--probe route to observing the predicted modulation.
\end{abstract}

\maketitle

\paragraph{Introduction.}

Vacuum pair production in strong electromagnetic fields---the Schwinger mechanism---is a paradigmatic nonperturbative prediction of quantum electrodynamics (QED), connecting tunneling physics to vacuum instability near the critical field scale. Seminal work established the effect for static fields~\cite{sauter_1931,heisenberg_euler,weisskopf,schwinger_1951}. Despite its conceptual simplicity, Schwinger pair creation remains a benchmark problem for strong-field quantum theory, with relevance to ultraintense-laser physics and related intense-field settings~\cite{dipiazza_rmp_2012,xie_review_2017,gonoskov_2022,fedotov_review,popruzhenko_ufn_2023}.

A striking outcome of many studies is the appearance of oscillatory ``fringe'' patterns in momentum-resolved spectra for \emph{structured, time-dependent} pulses~\cite{hebenstreit_prl_2009,dumlu_dunne_prl_2010,dumlu_dunne_prd_2011_1,dumlu_dunne_prd_2011_2,akkermans_prl_2012,abdukerim_plb_2013,kohlfuerst_prd_2013,hebenstreit_plb_2014,linder_prd_2015,aleksandrov_prd_2017}. These fringes admit a natural time-domain multiple-slit interpretation. In particular, semiclassical analyses based on complex-time turning points (or worldline instantons) show that fringes typically require multiple comparable saddle contributions, often generated by subcycle structure, pulse trains, or multi-component fields~\cite{dumlu_dunne_prl_2010,dumlu_dunne_prd_2011_1,dumlu_dunne_prd_2011_2,akkermans_prl_2012}.

We find that a single-lobe, carrier-free pulse can produce pronounced fringes purely from saddle competition, even when the real-time field contains \emph{no subcycle structure}. We compare two bell-shaped pulses with nearly identical real-time profiles, $e(z)=\mathrm{e}^{-z^2}$ and $e(z)=\mathrm{e}^{-z^2-z^4}$, and find a sharp change in the longitudinal spectrum in the nonadiabatic crossover. For the Gaussian, the spectra remain smooth and single-peaked, while for the deformed pulse they develop strong oscillatory fringes once the Keldysh parameter $\gamma$ approaches $\mathcal{O}(1)$. Exact numerical calculations (for both scalar and spinor QED) agree with a semiclassical turning-point (TP) analysis and identify the origin as a TP dominance transition: as $\gamma$ is tuned through unity, the dominant complex-time TP moves rapidly toward $i\infty$, so the next pair(s) become comparable and robust interference turns on. Crucially, this ``interference without subcycles'' is controlled by the \emph{global} analytic continuation of the pulse into complex time rather than by any real-time oscillations.

To connect to near-term experiments, we analyze the same mechanism in a gapped two-dimensional Dirac model relevant to graphene. Interband electron--hole generation in Dirac materials under strong in-plane electric fields is widely discussed as a condensed-matter analog of Schwinger pair production (or Landau--Zener tunneling), with both early and more recent theoretical treatments and proposed signatures in transport and (in some settings) radiation~\cite{allor_prd_2008,dora_prb_2010,lewkowicz_prb_2011,gavrilov_prd_2012,akal_prd_2016,akal_prd_2019,linder_prb_2018}. Epitaxial graphene on SiC provides a mature, wafer-scale platform in which substrate and buffer-layer coupling can break sublattice symmetry and generate an effective Dirac mass (often reported as a gap-like feature near the Dirac point)~\cite{zhou_nat-mat_2007,kim_prl_2008,enderlein_njp_2010,yu_apl_2013}. The gap is interface-dependent; intercalation can decouple the buffer layer and yield quasi-free-standing (essentially gapless) graphene~\cite{riedl_prl_2009}, offering a closely related reference system. Epitaxial graphene on SiC is routinely studied with broadband THz-to-mid-IR ultrafast spectroscopy and device-level in-plane driving~\cite{santos_sci-rep_2016,bianco_apl_2015,singh_jpd_2024,paschke_aqt_2020}. A natural primary readout in this setting is the pump-induced change of the low-energy optical/THz response (transient conductivity/absorption), which is routinely measured in ultrafast spectroscopy on SiC-supported graphene. Nonlinear THz response and emission channels in graphene are well documented~\cite{hafez_opt-mat_2020}, including THz-driven optical emission interpreted via field-induced $e$--$h$ generation~\cite{oladyshkin_prb_2017}. In the parameter regime emphasized below we focus on mid-IR/near-IR driving with duration $\tau\sim 10$--$100~\text{fs}$ and peak in-plane fields of order tens of~kV/cm, which spans the nonadiabatic crossover for experimentally relevant gaps. We also briefly comment on aligned graphene/hBN as an alternative route to a gapped Dirac spectrum via inversion-symmetry breaking and moir\'e potentials~\cite{wallbank_prb_2013_1,wallbank_prb_2013_2,wallbank_adp_2015,jung_prb_2017,li_comm-phys_2020}.

We establish the effect in QED and then demonstrate its graphene analog.

\paragraph{Spectral fringes in setups without subcycles.}

\begin{figure}[t]
\center{\includegraphics[width=0.98\linewidth]{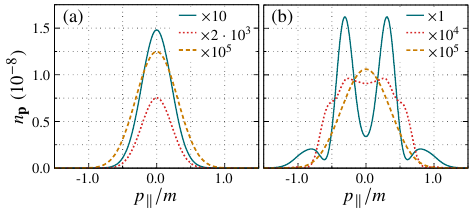}}
\caption{Momentum distributions of the produced particles as a function of longitudinal momentum $p_\parallel$ for $p_\perp=0$ in scalar QED for (a) a Gaussian pulse and (b) a deformed pulse. The numerical results are obtained for a peak field $E_0=0.1E_\text{c}$, where $E_\text{c}=m^2/|e|$ is the critical field strength. Pulse durations are $\tau=5m^{-1}$ (solid), $\tau=10m^{-1}$ (dotted), and $\tau=20m^{-1}$ (dashed), corresponding to $\gamma=2.0$, $1.0$, $0.5$, respectively. For clarity, curves are rescaled by the factors indicated in the panels.}
\label{fig:n_qed}
\end{figure}

We start with standard $(3+1)$ QED in units $\hbar=c=1$ and consider spatially homogeneous, linearly polarized electric fields
\begin{equation}
E(t)=E_0 e(t/\tau),
\end{equation}
where $e(z)$ is a carrier-free single-lobe profile with $e(0)=1$ and $\tau$ sets its duration. The nonadiabaticity is quantified by the Keldysh parameter $\gamma\equiv m/(|e|E_0\tau)$ (with $m$ the electron mass and $e<0$ its charge). We compare two smooth profiles,
$e(z)=\mathrm{e}^{-z^2}$ and $e(z)=\mathrm{e}^{-z^2-z^4}$, which are nearly indistinguishable on the real-time axis. In temporal gauge, $A(t)=-\int_{0}^{t} \! E(t')dt'$, we compute the asymptotic particle spectrum $n_{\mathbf{p}}$ by exact mode evolution (see, e.g., Refs.~\cite{aleksandrov_prd_2017,aleksandrov_prd_2016} for numerical details). In the adiabatic regime $\gamma\ll 1$, both pulses yield the expected smooth, bell-shaped longitudinal spectrum. In contrast, for $\gamma\sim 1$--$2$ the deformed pulse produces pronounced oscillations in the longitudinal momentum distribution, whereas the Gaussian spectrum remains smooth (Fig.~\ref{fig:n_qed}). We verify that this change is not a numerical artifact by comparing with a semiclassical WKB/TP construction and by repeating the analysis in both scalar and spinor QED.

\paragraph{Semiclassical mechanism.}

\begin{figure}[t]
\center{\includegraphics[width=0.98\linewidth]{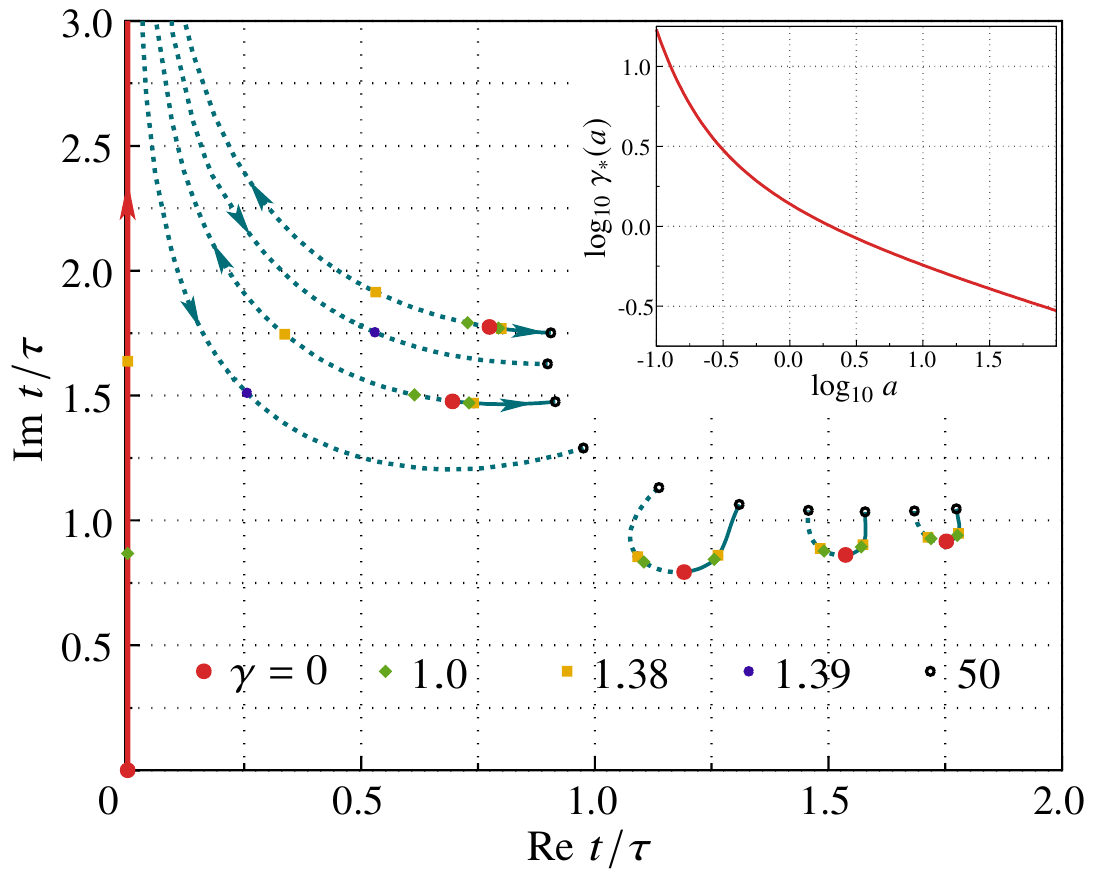}}
\caption{Turning-point trajectories for $\mathbf{p}=0$ in the first quadrant as $\gamma$ is varied (deformed pulse). For $\gamma\ll 1$, the dominant semiclassical contribution comes from the TP on the imaginary axis (red trajectory) that lies closest to the real axis. As $\gamma$ approaches $\gamma_*\simeq 1.381$, this TP rapidly migrates to $i\infty$ and becomes irrelevant, so that several remaining TPs acquire comparable weights; their interference produces pronounced fringes in the momentum spectrum. For clarity, only the 11 TPs closest to the origin are shown. The 10 initially subleading TPs form degenerate pairs at $\gamma=0$ and, for $\gamma\gg 1$, asymptote to the positions indicated by black circles. The full TP set is symmetric under reflection about the imaginary axis. Inset: critical value $\gamma_*(a)$ defined in Eq.~\eqref{eq:gammaStar_a} as a function of the deformation parameter $a$ for the pulse family~\eqref{eq:ea}.}
\label{fig:TP}
\end{figure}

The semiclassical interpretation follows the complex-time TP framework~\cite{dumlu_dunne_prd_2011_1,linder_prd_2015,popov_2005,oertel_prd_2019,taya_jhep_2021}: particle production is controlled by
zeros of the effective energy
\begin{equation}
Q_{\mathbf{p}}(t) = \sqrt{\pi_\perp^2 + [p_\parallel - e A (t)]^2}, \quad \pi_\perp \equiv \sqrt{m^2 + p_\perp^2},
\end{equation}
in complex time. Fringes arise when several TPs have comparable weight and interfere. The TPs $t_i$ are identified via $Q_{\mathbf{p}}(t_i) = 0$, which is equivalent to
\begin{equation}
a(t_i/\tau) = \frac{\pm i \pi_\perp - p_\parallel}{m}\, \gamma,
\end{equation}
where $a(z) = -\int_0^z \! e(z')dz'$ is the dimensionless vector potential. For the Gaussian pulse, one TP remains closest to the real axis for all $\gamma$, so a single saddle dominates. For the deformed pulse, the analytic continuation along the imaginary axis is qualitatively different: as shown in Fig.~\ref{fig:TP}, the would-be dominant TP rapidly loses relevance (effectively migrating to $i\infty$) near the critical value $\gamma_* = \int_{0}^{\infty}\! \mathrm{e}^{z^{2}-z^{4}} dz \approx 1.381$ (see Supplemental Material~\cite{SM} for a quantitative semiclassical analysis). In the crossover region $\gamma\sim \gamma_*$, several subleading TPs acquire comparable weights and interfere, producing fringes despite the absence of subcycle structure in $E(t)$. This \emph{turning-point dominance transition} provides a compact diagnostic---whether $\int_{0}^{\infty} E(i t) dt$ converges---indicating when a smooth single-lobe pulse can generate fringed spectra. In other words, the effect is set by the \emph{global} real-time pulse profile (through its analytic continuation), not by any oscillatory features on the real-time axis.

The TP dominance transition is not restricted to the specific deformation $e(z)=\mathrm{e}^{-z^{2}-z^{4}}$ but persists, for example, for a broad one-parameter family of smooth single-lobe pulses,
\begin{equation}
e_a(z)=\mathrm{e}^{-z^{2}-a z^{4}},\qquad a>0,
\label{eq:ea}
\end{equation}
which continuously interpolates between a nearly Gaussian envelope and a more rapidly decaying profile along the imaginary axis. In this family the critical value for the disappearance of the would-be leading TP is set by the convergent integral
\begin{equation}
\gamma_*(a)\equiv \int\limits _{0}^{\infty}\mathrm{e}^{z^{2}-a z^{4}} dz,
\label{eq:gammaStar_a}
\end{equation}
so that the onset of multi-saddle competition and spectral fringes is expected for $\gamma\sim\gamma_*(a)$ (see inset of Fig.~\ref{fig:TP}). Varying $a$ therefore provides a simple control knob that shifts the transition and can, e.g., move the fringe-onset threshold into a deeper tunneling regime (smaller $\gamma_*$), while leaving the real-time waveform single-lobed and carrier-free. Additional numerical examples for $a\neq 1$ are given in the Supplemental Material~\cite{SM}. Moreover, we have verified numerically that the same TP dominance transition occurs for super-Gaussian single-lobe profiles of the form $e(z)=\mathrm{exp}(-z^{4n})$ with $n=1$, $2$, $\ldots$, with the corresponding critical values $\gamma_*\approx 0.91$, $0.94$, $\ldots$ for these cases. Together with the deformation family~\eqref{eq:ea}, this confirms that the mechanism is not tied to a particular analytic form, but is generic across a broad class of smooth carrier-free pulses.

We now turn to a solid-state realization, showing the same TP dominance transition in gapped Dirac graphene under experimentally relevant ultrafast driving.

\paragraph{Condensed-matter analog in gapped Dirac graphene.}

To connect the TP dominance transition to near-term experiments, we consider a solid-state Schwinger analog in a two-dimensional massive Dirac system. A minimal model for gapped graphene in a spatially uniform, time-dependent in-plane field is
\begin{equation}
H_\mathbf{k} (t) = v_{\text{F}}\,\boldsymbol{\sigma}\!\cdot\!\bigl[\hbar\mathbf{k}-e\mathbf{A}(t)\bigr] + \Delta \sigma_z,
\label{eq:dirac_ham}
\end{equation}
with instantaneous band energy $\varepsilon_{\mathbf{k}}(t)=\sqrt{\Delta^2+ v_{\text{F}}^2 |\hbar\mathbf{k}-e\mathbf{A}(t)|^2}$. As in QED, TPs are given by zeros of the instantaneous gap, $\varepsilon_{\mathbf{k}}(t)=0$, so fringes arise from competition among complex saddles. We take $v_\text{F} = 10^6~\text{m}/\text{s}$ and label states by wave vector $\mathbf{k}$ (with momentum $\hbar \mathbf{k}$). In graphene-like Dirac systems, interband electron--hole generation in a uniform electric field is a well-known condensed-matter analog of Schwinger pair production~\cite{allor_prd_2008,dora_prb_2010,gavrilov_prd_2012,akal_prd_2016}. Starting from a filled valence band, the post-pulse conduction-band population $n_\mathbf{k}$ directly parallels the QED momentum spectrum. We quantify adiabaticity by comparing the intrinsic momentum scale $\Delta/v_{\text{F}}$ to the field impulse $|e|E_0\tau$, which defines the Dirac Keldysh parameter
\begin{equation}
\gamma_\text{D} = \frac{\Delta}{|e| v_{\text{F}} E_0 \tau}\,.
\label{eq:gamma_dirac}
\end{equation}
We compute $n_\mathbf{k}$ by solving the graphene quantum kinetic equations (QKEs) derived from the adiabatic-band (Bogoliubov) system; the full derivation for general and linearly polarized fields are given in the Supplemental Material~\cite{SM}, together with a constant-field benchmark that reproduces the Landau--Zener/Schwinger exponential in the quasistatic limit, $n_{\mathbf{k}} \approx \mathrm{exp}(-\pi E_\text{D}/E_0)$, where $E_\text{D} = \Delta^2/(|e|\hbar v_\text{F})$ is the critical field strength~\cite{dora_prb_2010,gavrilov_prd_2012,akal_prd_2016,akal_prd_2019}. For gapless graphene ($\Delta = 0$), the QKE system was obtained in Refs.~\cite{smolyansky_particles_2019,smolyansky_particles_2020}.

\begin{figure}[t]
\center{\includegraphics[width=0.98\linewidth]{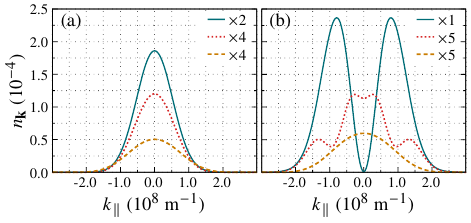}}
\caption{Longitudinal momentum distributions of the produced carriers in gapped graphene as a function of $k_\parallel$ for $k_\perp=0$ and $\Delta=0.1~\text{eV}$ for (a) Gaussian and (b) deformed pulses. The field amplitude is $E_0=40~\text{kV}/\text{cm}$. Pulse durations are $\tau=20~\text{fs}$ (solid), $\tau=30~\text{fs}$ (dotted), and $\tau=50~\text{fs}$ (dashed), corresponding to $\gamma_{\text{D}}=1.25$, $0.833$, and $0.5$, respectively. For clarity, curves are rescaled by the factors indicated in the panels.}
\label{fig:graphene}
\end{figure}

Figure~\ref{fig:graphene} shows that the QED pattern carries over to gapped graphene: for a Gaussian envelope the longitudinal spectra remain smooth, while for the deformed pulse they develop pronounced fringes once $\gamma_\text{D}$ approaches unity. For $\gamma_\text{D}\ll 1$, both envelopes yield spectra whose peak values are close to the constant-field estimate, consistent with a quasistatic nonperturbative production channel. In Fig.~\ref{fig:graphene} we used $E_0 = 40~\text{kV}/\text{cm}$ and $\Delta=0.1~\text{eV}$, leading to $E_\text{D} \approx 152~\text{kV}/\text{cm}$. In all graphene calculations, the populated wave vectors satisfy $|\mathbf{k}|\ll a^{-1}\simeq 4 \times 10^{9}~\text{m}^{-1}$ (with lattice constant $a=2.46~\si{\angstrom}$), so the low-energy gapped Dirac model remains well justified.

To gauge the expected signal, it is useful to translate the post-pulse carrier distribution into experimentally accessible observables. While our microscopic output is the post-pulse conduction-band occupation in momentum space, the most direct observables are pump-induced changes in THz-to-mid-IR conductivity/absorption/reflectivity, which probe momentum-weighted integrals of the distribution. An order-of-magnitude estimate for the total number of excited carriers $N$ in the illuminated region is obtained by integrating the distribution over the characteristic momentum scale $k_\Delta\equiv \Delta/(\hbar v_\text{F})\sim 10^{8}~\text{m}^{-1}$ (for $\Delta\simeq 0.1~\text{eV}$). Including spin and valley degeneracy ($g=4$), one finds $N \sim g f_\text{max} (k_\Delta L)^2/(4\pi)$, where $f_\text{max}$ is the peak occupation and $L$ is the characteristic interaction length. For a far-field spot of linear size $L\sim 10$--$100~\qty{}{\micro\metre}$, this corresponds to a non-negligible number of photoexcited carriers per pulse even for $f_\text{max} \sim 10^{-8}$, implying a measurable pump-induced modification of the low-energy response (conductivity/absorption) in standard ultrafast infrared/THz probes.

\begin{figure}[t]
\center{\includegraphics[width=0.99\linewidth]{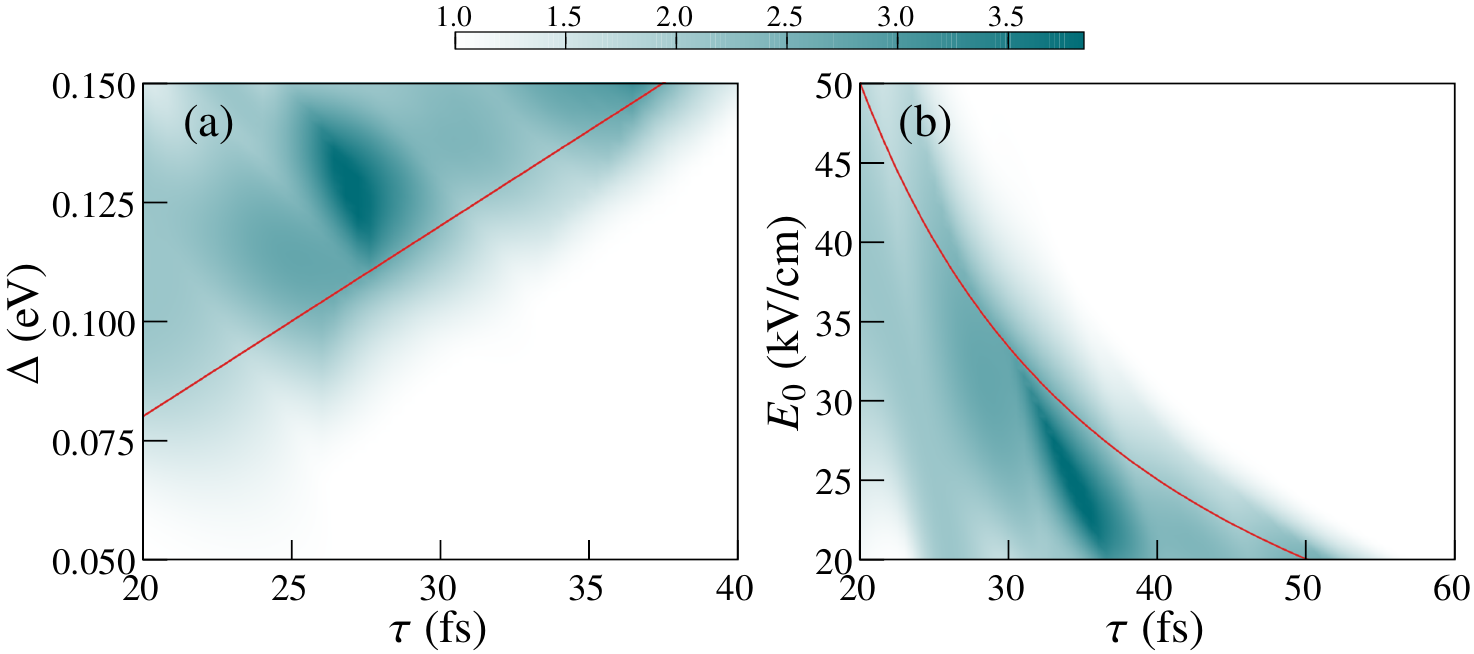}}
\caption{Dimensionless fringe-visibility measure $\nu$ for the deformed single-lobe pulse $e(z) = \mathrm{exp}(-z^2-z^4)$ in the gapped-graphene Dirac model. Panel (a) shows $\nu$ in the $(\tau,\Delta)$ plane at fixed field amplitude $E_0 = 40~\text{kV}/\text{cm}$. Panel (b) displays $\nu$ in the $(\tau, E_0)$ plane at fixed gap parameter $\Delta=0.1~\text{eV}$. The red curves mark the nonadiabatic crossover $\gamma_{\text{D}}=1.0$, along which the onset of pronounced oscillatory fringes is observed.}
\label{fig:heatmaps}
\end{figure}

To quantify the strength of spectral fringes we introduce a simple extrema-based ``wiggliness'' measure. For each longitudinal momentum distribution, we identify the sequence of successive local maxima and minima (including the endpoints of the plotted interval) and sum the \emph{absolute changes} of the spectrum between neighboring entries. Normalizing this total variation by the peak value yields a dimensionless fringe-visibility measure $\nu$: a smooth, single-peaked spectrum gives $\nu\simeq 1$, while resolved oscillations increase $\nu>1$. Using $\nu$ as a compact diagnostic, Fig.~\ref{fig:heatmaps} summarizes where the interference sets in for the deformed single-lobe pulse. Panel~(a) shows $\nu$ in the $(\tau,\Delta)$ plane at fixed $E_0=40~\text{kV}/\text{cm}$, and panel~(b) shows $\nu$ in the $(\tau,E_0)$ plane at fixed $\Delta=0.1~\text{eV}$. In both cases the enhanced-fringe region tracks the nonadiabatic crossover $\gamma_\text{D} \simeq 1$, consistent with the TP dominance transition picture.

Experimentally, the fringe-onset maps indicate where the momentum distribution is most sensitive to pulse shape and therefore where the largest pump-induced changes in low-energy optical response are expected. In far-field pump--probe measurements this sensitivity can be accessed through transient absorption/reflectivity in the THz-to-mid-IR range, which probes momentum-shell averages set by the interband resonance condition $2\varepsilon_{\mathbf{k}}\simeq \hbar\Omega$ and is therefore naturally sensitive to structured post-pulse populations~\cite{dawlaty_apl_2008,horng_prb_2011,jnawali_nano_lett_2013}. In particular, after angular averaging the post-pulse distribution can be recast as an energy-resolved spectrum $dN/d\varepsilon$; in the parameter range of Fig.~\ref{fig:heatmaps} this spectrum exhibits oscillations with characteristic spacing $\delta\varepsilon_\text{fr}\sim 20~\text{meV}$ (see Supplemental Material~\cite{SM}). Since interband probes address $2\varepsilon$ through $\hbar\Omega$, detecting this modulation requires a spectral resolution on the order of $\Delta(\hbar\Omega)\lesssim 2 \delta\varepsilon_\text{fr}\sim 40~\text{meV}$, which is feasible with narrowband ultrafast probes. In device geometries, the same nonequilibrium carrier population can also generate transient currents and radiation signals whose connection to field-driven Landau--Zener/Schwinger carrier creation has been discussed in the graphene literature~\cite{dora_prb_2010,lewkowicz_prb_2011,gavrilov_prd_2012,gavrilov_universe_2020,oladyshkin_prb_2017}. Finally, note that finite spectral/momentum resolution and dephasing primarily reduce fringe contrast; the energy-domain analysis in the Supplemental Material shows the modulation survives realistic broadening scales.

A possible concern is that an idealized strictly single-lobe field has a nonvanishing time integral, whereas a freely propagating far-field pulse cannot carry a static (zero-frequency) component. In practice, waveform-asymmetric single-/half-cycle transients are accompanied by a weak, long tail of opposite sign that restores zero net field area, and similar effective single-lobe driving can also be implemented through near-field coupling from antennas or circuit elements~\cite{schmuttenmaer_2004,isgandarov_2021}. In the Supplemental Material~\cite{SM}, we explicitly verify that the fringe formation persists for zero-area pulses obtained by adding such a compensating tail.

\paragraph{Discussion and outlook.}

Our results show that nonperturbative spectra can depend qualitatively on the global analytic structure of otherwise smooth, carrier-free pulses. In QED this yields ``interference without subcycles'' driven by turning-point dominance transitions rather than real-time carrier oscillations, and we demonstrate the same mechanism in a realistic solid-state analog. A natural question is whether this sensitivity can be controlled experimentally. In practice, experiments shape only the real-time waveform; the turning-point picture provides an organizing principle that predicts when small, smooth deformations of a single-lobe pulse trigger a dominance switch among saddles and hence qualitative spectral changes. Importantly, the effect is robust across broad families of smooth single-lobe profiles, so it does not rely on fine tuning. Conversely, this sensitivity also suggests a nonperturbative pulse diagnostic: fringe onset and evolution can fingerprint the effective field experienced by the carriers, complementing conventional waveform characterization.

Epitaxial graphene naturally corresponds to far-field pump--probe measurements on an extended sheet, where the relevant interaction length $L$ is set by the smaller of the illuminated spot size and the device dimension along the field direction. A kinematic estimate for ballistic escape is $t_\text{esc}\sim L/v_{\text{F}}$~\cite{dora_prb_2010}. For typical far-field spots $L\sim 10$--$100~\qty{}{\micro\metre}$ one finds $t_\text{esc}\sim 10$--$100~\text{ps}$, so for the few-cycle mid-IR/near-IR driving emphasized here ($\tau\sim 10$--$100~\text{fs}$) the creation stage satisfies $\tau\ll t_\text{esc}$ and transport out of the driven region is negligible. Relaxation and dephasing primarily add broadening that reduces fringe contrast, while leaving the dominance-transition onset scale unchanged.

\begin{acknowledgments}
We are grateful to Prof.~I.~S.~Terekhov for valuable discussions. The study was funded by the Foundation for the Advancement of Theoretical Physics and Mathematics BASIS (Project No.~25-1-3-48-1). The work of M.~A.~Dorodnyi was performed at the Saint Petersburg Leonhard Euler International Mathematical Institute and supported by the Ministry of Science and Higher Education of the Russian Federation (agreement No.~075–15–2025–343).
\end{acknowledgments}



\clearpage
\onecolumngrid
\appendix
\pagenumbering{roman}
\setcounter{page}{1}
\setcounter{equation}{0}
\setcounter{figure}{0}
\setcounter{table}{0}
\renewcommand{\theequation}{S\arabic{equation}}
\renewcommand{\thefigure}{S\arabic{figure}} 

\begin{center}
{\bf \large Spectral fringes without subcycles in Schwinger pair production and Dirac materials}

\vspace{0.2cm}

{I.~A.~Aleksandrov, M.~A.~Dorodnyi, and E.~D.~Akimkina}

\vspace{0.1cm}

{\it Supplemental Material}

\end{center}

This Supplemental Material is organized as follows. Appendix~A summarizes the semiclassical turning-point framework (including the Stokes/prefactor conventions) used to interpret the spectra in scalar and spinor QED. Appendix~B contains several additional examples of the particle momentum distributions in the case of deformed pulses $e_a(z) = \mathrm{exp}(-z^2 - az^4)$. Appendix~C derives the quantum kinetic equations for the gapped Dirac model and records the coefficient functions used in our numerics.

\vspace{0.1cm}

\begin{center}
{\bf A. WKB APPROXIMATION IN SCALAR AND SPINOR QED}
\end{center}

\vspace{0.1cm}

Here we briefly recap the main WKB expressions which allow one to approximately evaluate the momentum spectra $n_\mathbf{p}$. In the case of a linearly polarized field $E(t) = -\dot{A} (t)$, we define the adiabatic energy
\begin{equation}
Q_{\mathbf{p}}(t) = \sqrt{\pi_\perp^2 + [p_\parallel - e A (t)]^2},
\end{equation}
where $\pi_\perp = \sqrt{m^2 + p_\perp^2}$. Assuming that the vector potential is a regular function of a complex variable $t$, we identify the upper-half-plane turning points (TPs) for given $\mathbf{p}$:
\begin{equation}
Q_{\mathbf{p}}(t_i) = 0.
\end{equation}
The complex conjugates $t_i^*$ are also TPs. We then introduce
\begin{equation}
\rho_i (t) = \int \limits_{t_i}^{t} Q_{\mathbf{p}}(t')dt'.
\end{equation}
It is clear that $\rho_i (t_i) = 0$. The Stokes lines are defined as the following sets:
\begin{equation}
\mathcal{S}_i = \big \{ t \in \mathbb{C}: \mathrm{Re} \, \rho_i (t) = 0 \big \}.
\label{eq:stokes_def}
\end{equation}
It turns out that $t_i$ and $t_i^*$ are connected by the corresponding Stokes line $\mathcal{S}_i$~\cite{taya_jhep_2021}. The intersection of $\mathcal{S}_i$ and the real axis is denoted by $s_i$. According to Refs.~\cite{dumlu_dunne_prd_2011_1,taya_jhep_2021}, in scalar QED the momentum distributions can be obtained by the following approximate recipe:
\begin{equation}
n_\mathbf{p} \approx \sum_{i} \mathrm{e}^{-2 \sigma_{\mathbf{p}}^{(i)}} + 2 \sum_{i<j} \cos [2 \theta_{\mathbf{p}}^{(ij)} ] \mathrm{e}^{-\sigma_{\mathbf{p}}^{(i)}-\sigma_{\mathbf{p}}^{(j)}},
\label{eq:np_WKB}
\end{equation}
where we have introduced {\it real-valued} functions
\begin{equation}
\theta_{\mathbf{p}}^{(ij)} \equiv \int \limits_{s_i}^{s_j}Q_{\mathbf{p}}(t)dt, \qquad \sigma_{\mathbf{p}}^{(i)} \equiv -i \int \limits_{t_i^*}^{t_i}Q_{\mathbf{p}}(t)dt.
\end{equation}
If there are multiple TPs with relatively large exponential factors $\mathrm{exp} (-\sigma_{\mathbf{p}}^{(i)})$, then the second term in Eq.~\eqref{eq:np_WKB} gives rise to nontrivial interference patterns. In spinor QED, the final WKB prediction reads~\cite{dumlu_dunne_prd_2011_1}
\begin{equation}
n_\mathbf{p} \approx \sum_{i} \mathrm{e}^{-2 \sigma_{\mathbf{p}}^{(i)}} + 2 \sum_{i<j} (-1)^{i-j} \cos [2 \theta_{\mathbf{p}}^{(ij)} ] \mathrm{e}^{-\sigma_{\mathbf{p}}^{(i)}-\sigma_{\mathbf{p}}^{(j)}}.
\label{eq:np_WKB_spinor}
\end{equation}

In order to identify the TPs, it is convenient to represent the external field in the form
\begin{equation}
E(t) = E_0 e(t/\tau),
\end{equation}
where one has to specify the field amplitude $E_0$, its duration $\tau$, and the profile $e(z)$. We assume $e(0)=1$. After introducing
\begin{equation}
a(z) = -\int \limits_0^z \! e(z') dz',
\end{equation}
we can choose the vector potential in the form
\begin{equation}
A(t) = E_0 \tau a (t/\tau).
\end{equation}
The TPs $t_i$ can be found via 
\begin{equation}
a(t_i/\tau) = \frac{\pm i \pi_\perp - p_\parallel}{m}\, \gamma,
\label{eq:TP_eq_app}
\end{equation}
where
\begin{equation}
\gamma = \frac{m}{|eE_0|\tau}
\end{equation}
is the Keldysh parameter. By solving Eq.~\eqref{eq:TP_eq_app}, we find the TPs for various values of $\gamma$. The crucial point here is that the motion of the TPs becomes highly nontrivial in the case of a deformed Gaussian profile $e(z) = \mathrm{e}^{-z^2-z^4}$. As was demonstrated in the main text, at sufficiently large $\gamma$, several TPs simultaneously provide relatively large contributions, so according to Eqs.~\eqref{eq:np_WKB} and \eqref{eq:np_WKB_spinor}, the particle spectra gain interference fringes.

\vspace{0.2cm}

\begin{center}
{\bf B. FRINGED SPECTRA FOR DEFORMED PULSES $e_a(z) = \mathrm{e}^{-z^2-az^4}$}
\end{center}

\vspace{0.1cm}

\begin{figure}[h]
\center{\includegraphics[width=0.6\linewidth]{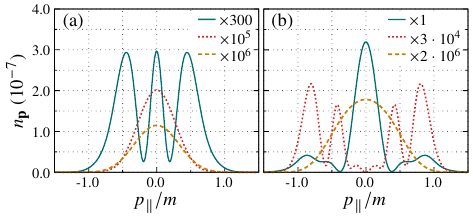}}
\caption{Momentum distributions of the produced particles as a function of longitudinal momentum $p_\parallel$ for $p_\perp=0$ in scalar QED for a deformed pulse~\eqref{eq:SM_ea}  with (a) $a = 0.5$ and (b) $a = 2.0$. The numerical results are obtained for a peak field $E_0=0.1E_\text{c}$, where $E_\text{c}=m^2/|e|$ is the critical field strength. Pulse durations are $\tau=5m^{-1}$ (solid), $\tau=10m^{-1}$ (dotted), and $\tau=20m^{-1}$ (dashed), leading to $\gamma=2.0$, $1.0$, $0.5$, respectively. For clarity, the plotted number densities are rescaled by the multiplicative factors indicated in the panels.}
\label{fig:SM_ea}
\end{figure}

To support our discussion of the single-lobe pulses
\begin{equation}
e_a(z)=\mathrm{e}^{-z^{2}-a z^{4}} \qquad (a>0),
\label{eq:SM_ea}
\end{equation}
we provide here additional numerical examples of the momentum distributions in scalar QED. We compare two pulses with $a=0.5$ and $a=2.0$, respectively. The critical values of the Keldysh parameter~\eqref{eq:gammaStar_a} amount to $\gamma_* (0.5) \approx 2.08$ and $\gamma_* (2.0) \approx 1.01$. The example spectra are displayed in Fig.~\ref{fig:SM_ea}. As anticipated from the $\gamma_*(a)$ indicator, the onset of pronounced fringes follows the location of the turning-point dominance transition. For $a=0.5$ (panel~a), the spectrum at $\gamma\simeq2$ already lies close to $\gamma_*(0.5)\approx2.08$ and exhibits clear oscillatory modulation, while moving deeper into the adiabatic regime ($\gamma=1,0.5$) suppresses the fringes toward a smooth single-peaked shape. For the more strongly deformed profile $a=2.0$ (panel~b), the critical value is much smaller, $\gamma_*(2.0)\approx1.01$, and correspondingly the qualitative change occurs already around $\gamma\simeq1$: spectra above and below this threshold display markedly different fringe visibility. These additional examples therefore corroborate the $\gamma_*(a)$-based diagnostic and demonstrate that the semiclassical mechanism is robust across the deformation family~\eqref{eq:SM_ea}, rather than being specific to the single case $a=1$.

\newpage

\begin{center}
{\bf C. DERIVATION OF THE QKE SYSTEM IN GAPPED GRAPHENE}
\end{center}

\vspace{0.1cm}

Here we present a derivation of the QKE system in a massive ($2+1$) Dirac model. The one-particle Hamiltonian in the presence of a time-dependent electric field is
\begin{equation}
  \mathcal{H}(t)= v_\text{F} \, \boldsymbol{\sigma} \cdot [- i \hbar \boldsymbol{\nabla} - e \mathbf{A} (t) ] + \Delta \sigma_z,\qquad \mathbf{E}(t)=-\dot{\mathbf{A}}(t),
\end{equation}
where $e<0$ is the electron charge, $v_\text{F}$ is the Fermi velocity, and $\Delta$ the Dirac mass term (half the band gap), $\Delta=E_\text{g}/2$. We assume that the electric field vanishes unless $t_\text{in} < t < t_\text{out}$. We define the following spinors:
\begin{equation}
  u_{\mathbf{p}}
  =\frac{1}{\sqrt{2\hbar\omega_{\mathbf{p}}}}
    \begin{pmatrix}
      \sqrt{\hbar\omega_{\mathbf{p}}+\Delta}\\[4mm]
      \dfrac{p_x+i p_y}{|\mathbf{p}|}\sqrt{\hbar\omega_{\mathbf{p}}-\Delta}
    \end{pmatrix},\qquad
  v_{\mathbf{p}}
  =\frac{1}{\sqrt{2\hbar\omega_{\mathbf{p}}}}
    \begin{pmatrix}
      -\,\dfrac{p_x-i p_y}{|\mathbf{p}|}\sqrt{\hbar\omega_{\mathbf{p}}-\Delta}\\[4mm]
      \sqrt{\hbar\omega_{\mathbf{p}}+\Delta}
    \end{pmatrix},
  \label{eq:uv_spinor_bb}
\end{equation}
where
\begin{equation}
\hbar \omega_{\mathbf{p}}
  =\sqrt{\Delta^2+v_\text{F}^2|\mathbf{p}|^2}.
\end{equation}
They satisfy the orthonormality and completeness relations
\begin{equation}
  u_{\mathbf{p}}^\dagger u_{\mathbf{p}}=v_{\mathbf{p}}^\dagger v_{\mathbf{p}}=1,\qquad
  u_{\mathbf{p}}^\dagger v_{\mathbf{p}}=0,\qquad
u_{\mathbf{p}}u_{\mathbf{p}}^\dagger+v_{\mathbf{p}}v_{\mathbf{p}}^\dagger=1.
\end{equation}
We also have
\begin{equation}
(v_\text{F}\,\boldsymbol{\sigma} \mathbf{p} + \Delta \sigma_z) u_{\mathbf{p}} = \hbar \omega_{\mathbf{p}} u_{\mathbf{p}},\qquad (v_\text{F}\,\boldsymbol{\sigma} \mathbf{p} + \Delta \sigma_z) v_{\mathbf{p}} = -\hbar \omega_{\mathbf{p}} v_{\mathbf{p}}.
\end{equation}
This allows us to define an adiabatic set of wave functions with positive and negative energies:
\begin{equation}
  \Phi^{(+)}_{\mathbf{p}}(\mathbf{x}; t)
  =\frac{1}{2\pi}\,\mathrm{e}^{i\mathbf{p}\mathbf{x}/\hbar}\,
    u_{\mathbf{p} - e \mathbf{A}(t)},
  \qquad
  \Phi^{(-)}_{\mathbf{p}}(\mathbf{x}; t)
  =\frac{1}{2\pi}\,\mathrm{e}^{-i\mathbf{p}\mathbf{x}/\hbar}\,
    v_{-\mathbf{p} - e \mathbf{A}(t)},
  \label{eq:phi_pm_def}
\end{equation}
so that
\begin{equation}
\mathcal{H}(t)\Phi^{(+)}_{\mathbf{p}} = \hbar \omega_{\mathbf{p} - e \mathbf{A}(t)} \Phi^{(+)}_{\mathbf{p}},
  \qquad
  \mathcal{H}(t)\Phi^{(-)}_{\mathbf{p}} = -\hbar \omega_{\mathbf{p} + e \mathbf{A}(t)} \Phi^{(-)}_{\mathbf{p}}.
\end{equation}
The momentum is connected with the wave vector via $\mathbf{p} = \hbar \mathbf{k}$.

The exact solution of the Dirac equation
$i\hbar\partial_t\Psi(\mathbf{x},t) = \mathcal{H}(t)\Psi (\mathbf{x},t)$ can be expanded in this adiabatic basis. Let us consider the following {\it in} solution, i.e., wave function fixed by its form at $t=t_\text{in}$:
\begin{equation}
  {}_- \Psi_{-\mathbf{p}}(\mathbf{x},t)
  =\alpha_{\mathbf{p}}(t)\,\Phi^{(-)}_{-\mathbf{p}}(\mathbf{x}; t)
  +\beta_{\mathbf{p}}(t)\,\Phi^{(+)}_{\mathbf{p}}(\mathbf{x}; t)
  \label{eq:psi_expansion_bb}
\end{equation}
with $\alpha_{\mathbf{p}}(t_\text{in})=1$, $\beta_{\mathbf{p}}(t_\text{in})=0$. The next step is to express the Dirac equation in terms of the coefficients $\alpha_{\mathbf{p}}(t)$ and $\beta_{\mathbf{p}}(t)$. We define $\mathbf{q}(t)=\mathbf{p}-e\mathbf{A}(t)$. Using $\dot{\mathbf{q}}(t)=e\mathbf{E}(t)$ and the chain rule
\begin{equation}
  \dot{u}_{\mathbf{q}(t)}=e \mathbf{E}(t) \cdot \boldsymbol{\nabla}_{\mathbf{p}}u_{\mathbf{p}} \big |_{\mathbf{p} = \mathbf{q}(t)},
  \qquad
  \dot{v}_{\mathbf{q}(t)}= e \mathbf{E}(t) \cdot \boldsymbol{\nabla}_{\mathbf{p}}v_{\mathbf{p}} \big |_{\mathbf{p} = \mathbf{q}(t)},
  \label{eq:udot_chainrule}
\end{equation}
we substitute Eq.~\eqref{eq:psi_expansion_bb} into the Dirac equation and project onto
$v_{\mathbf{q}}^\dagger$ and $u_{\mathbf{q}}^\dagger$.
This yields the exact coupled system
\begin{align}
  \dot{\alpha}_{\mathbf{p}}
  +\left(v_{\mathbf{q}}^\dagger \dot{v}_{\mathbf{q}}\right)\alpha_{\mathbf{p}}
  +\left(v_{\mathbf{q}}^\dagger \dot{u}_{\mathbf{q}}\right)\beta_{\mathbf{p}}
  &= i\omega_{\mathbf{q}} \alpha_{\mathbf{p}},
  \label{eq:alpha_raw}\\
  \dot{\beta}_{\mathbf{p}}
  +\left(u_{\mathbf{q}}^\dagger \dot{u}_{\mathbf{q}}\right)\beta_{\mathbf{p}}
  +\left(u_{\mathbf{q}}^\dagger \dot{v}_{\mathbf{q}}\right)\alpha_{\mathbf{p}}
  &=-i\omega_{\mathbf{q}} \beta_{\mathbf{p}},
  \label{eq:beta_raw}
\end{align}
where $\mathbf{q}\equiv \mathbf{q}(t)$ and $\omega_{\mathbf{q}}\equiv\omega_{\mathbf{q}(t)}$. These equations can be simplified. First we combine the following terms:
\begin{equation}
i\Omega^{(-)}_{\mathbf{p}} \equiv i\omega_{\mathbf{q}} - \left(v_{\mathbf{q}}^\dagger \dot{v}_{\mathbf{q}}\right),\qquad
i\Omega^{(+)}_{\mathbf{p}} \equiv i\omega_{\mathbf{q}} + \left(u_{\mathbf{q}}^\dagger \dot{u}_{\mathbf{q}}\right).
\end{equation}
Note that $\Omega^{(\pm)}_{\mathbf{q}}$ turn out to be real. We also define the interband coupling
\begin{equation}
W_{\mathbf{p}}(t)\equiv v_{\mathbf{q}}^\dagger \dot{u}_{\mathbf{q}} = e \mathbf{E}(t) \big [ v_{\mathbf{p}}^\dagger \boldsymbol{\nabla}_{\mathbf{p}}u_{\mathbf{p}} \big ] \Big |_{\mathbf{p} = \mathbf{q}(t)} .
\end{equation}
It follows that
\begin{equation}
u_{\mathbf{q}}^\dagger \dot{v}_{\mathbf{q}} = e \mathbf{E}(t) \big [ u_{\mathbf{p}}^\dagger \boldsymbol{\nabla}_{\mathbf{p}} v_{\mathbf{p}} \big ] \Big |_{\mathbf{p} = \mathbf{q}(t)} = -W_{\mathbf{p}}^*(t).
\end{equation}
Therefore, we obtain
\begin{align}
\dot{\alpha}_{\mathbf{p}} (t) &= i \Omega^{(-)}_{\mathbf{p}} (t) \alpha_{\mathbf{p}} (t) - W_{\mathbf{p}} (t) \beta_{\mathbf{p}} (t), \\
\dot{\beta}_{\mathbf{p}} (t) &= -i \Omega^{(+)}_{\mathbf{p}} (t) \beta_{\mathbf{p}} (t) + W^*_{\mathbf{p}} (t) \alpha_{\mathbf{p}} (t).
\end{align}
By absorbing the trivial phase factors
\begin{equation}
\alpha_{\mathbf{p}} (t) = \tilde{\alpha}_{\mathbf{p}} (t) \, \mathrm{exp} \Bigg [ i \int\limits_{t_\text{in}}^t \Omega^{(-)}_{\mathbf{p}} (t') dt' \Bigg ],\qquad
\beta_{\mathbf{p}} (t) = \tilde{\beta}_{\mathbf{p}} (t) \, \mathrm{exp} \Bigg [ -i \int\limits_{t_\text{in}}^t \Omega^{(+)}_{\mathbf{p}} (t') dt' \Bigg ],
\end{equation}
we arrive at the standard reduced Bogoliubov system
\begin{align}
\dot{\tilde{\alpha}}_{\mathbf{p}}(t) &= -W_{\mathbf{p}}(t)\, \mathrm{e}^{-2i\Theta_{\mathbf{p}}(t)}\, \tilde{\beta}_{\mathbf{p}}(t), \label{eq:system_alpha} \\
 \dot{\tilde{\beta}}_{\mathbf{p}}(t) &= W_{\mathbf{p}}^*(t)\,\mathrm{e}^{2i\Theta_{\mathbf{p}}(t)}\,\tilde{\alpha}_{\mathbf{p}}(t), \label{eq:system_beta}
\end{align}
where 
\begin{equation}
\Theta_{\mathbf{p}}(t)\equiv \int\limits_{t_\text{in}}^t \Omega_{\mathbf{p}}(t') dt', \qquad \Omega_{\mathbf{p}} (t) \equiv \frac{1}{2} \big [ \Omega^{(-)}_{\mathbf{p}} (t) + \Omega^{(+)}_{\mathbf{p}} (t) \big ].
\end{equation}
The initial conditions read $\tilde{\alpha}(t_\text{in})=1$, $\tilde{\beta} (t_\text{in})=0$.
The momentum-resolved carrier density as a function of the  wave vector $\mathbf{k}$ is then
\begin{equation}
n_{\mathbf{k}}=|\tilde{\beta}_{\hbar \mathbf{k}}(t_\text{out} )|^2.
\end{equation}
Note that for each $t$, we have $|\tilde{\alpha}_{\mathbf{p}}|^2 + |\tilde{\beta}_{\mathbf{p}}|^2 = 1$, which reflects the unitary evolution of the wave function.

Next we employ the explicit form of the spinors~\eqref{eq:uv_spinor_bb}. For $\dot{\Theta}_{\mathbf{p}} = \Omega_{\mathbf{p}}$ we obtain
\begin{equation}
\Omega_{\mathbf{p}} (t) = \omega_{\mathbf{q}} + \frac{e(\hbar \omega_{\mathbf{q}} - \Delta)(q_xE_y - q_y E_x)}{2\hbar \omega_{\mathbf{q}} |\mathbf{q}|^2}.
\end{equation}
Second, we introduce
\begin{equation}
W_{\mathbf{p}}(t) = \frac{\lambda_{1,\mathbf{p}}(t)+i \lambda_{2,\mathbf{p}}(t)}{2}
\end{equation}
and find
\begin{align}
\lambda_{1,\mathbf{p}} (t) &= \frac{ev_\text{F}}{\hbar \omega_{\mathbf{q}} |\mathbf{q}|^2} \bigg [ \frac{\Delta}{\hbar \omega_{\mathbf{q}}} q_x (\mathbf{q},\mathbf{E}) - q_y (q_x E_y - q_y E_x) \bigg ], \\
\lambda_{2,\mathbf{p}} (t) &= \frac{ev_\text{F}}{\hbar \omega_{\mathbf{q}} |\mathbf{q}|^2} \bigg [ \frac{\Delta}{\hbar \omega_{\mathbf{q}}} q_y (\mathbf{q},\mathbf{E}) + q_x (q_x E_y - q_y E_x) \bigg ].
\end{align}
The system~\eqref{eq:system_alpha}, \eqref{eq:system_beta} can be rewritten in terms of the real-valued functions. Let us introduce
\begin{equation}
  f_{\mathbf{p}}(t)=|\tilde{\beta}_{\mathbf{p}}(t)|^2,\qquad
  u_{\mathbf{p}}(t)=2\,\mathrm{Re} \left[\tilde{\alpha}_{\mathbf{p}}(t)\tilde{\beta}_{\mathbf{p}}^*(t)\mathrm{e}^{2i\Theta_{\mathbf{p}}(t)}\right],\qquad
  v_{\mathbf{p}}(t)=2\,\mathrm{Im} \left[\tilde{\alpha}_{\mathbf{p}}(t)\tilde{\beta}_{\mathbf{p}}^*(t)\mathrm{e}^{2i\Theta_{\mathbf{p}}(t)}\right].
\end{equation}
One straightforwardly obtains the necessary QKE system:
\begin{align}
  \dot f_{\mathbf{p}}(t) &= \tfrac12 \lambda_{1,\mathbf{p}}(t) u_{\mathbf{p}}(t)+\tfrac12 \lambda_{2,\mathbf{p}}(t) v_{\mathbf{p}}(t),
  \label{eq:qke_f}\\
  \dot u_{\mathbf{p}}(t) &= \lambda_{1,\mathbf{p}}(t)\big[1-2f_{\mathbf{p}}(t)\big]-2\,\Omega_{\mathbf{p}}(t) v_{\mathbf{p}}(t),
  \label{eq:qke_u}\\
  \dot v_{\mathbf{p}}(t) &= \lambda_{2,\mathbf{p}}(t)\big[1-2f_{\mathbf{p}}(t)\big]+2\,\Omega_{\mathbf{p}}(t) u_{\mathbf{p}}(t),
  \label{eq:qke_v}
\end{align}
with initial conditions $f_{\mathbf{p}}(t_\text{in})=u_{\mathbf{p}}(t_\text{in})=v_{\mathbf{p}}(t_\text{in})=0$. As a useful internal consistency check for both the derivation and numerics, the system admits the exact invariant
\begin{equation}
  \big[1-2f_{\mathbf{p}}(t)\big]^2+u^2_{\mathbf{p}}(t)+v^2_{\mathbf{p}}(t)=1,
\end{equation}
which should be preserved by the numerical integration up to solver tolerance. The wave-vector distribution is then given by
\begin{equation}
n_\mathbf{k} = f_{\hbar \mathbf{k}}(t_\text{out}).
\end{equation}
In the case of linear polarization $E_y = 0$, we define $q_x = q_\parallel (t) = p_\parallel - eA(t)$ and $q_y = q_\perp$. We then find
\begin{align}
\Omega_{\mathbf{p}} (t) &= \omega_{\mathbf{q}} - \frac{e E(t) q_\perp (\hbar \omega_{\mathbf{q}} - \Delta)}{2\hbar \omega_{\mathbf{q}} |\mathbf{q}|^2}, \\
\lambda_{1,\mathbf{p}} (t) &= \frac{eE(t)v_\text{F}}{\hbar \omega_{\mathbf{q}} |\mathbf{q}|^2} \bigg ( \frac{\Delta}{\hbar \omega_{\mathbf{q}}} q_\parallel^2(t) + q_\perp^2 \bigg ), \\
\lambda_{2,\mathbf{p}} (t) &= -\frac{eE(t) q_\parallel (t) q_\perp v_\text{F} (\hbar \omega_{\mathbf{q}} - \Delta)}{(\hbar \omega_{\mathbf{q}})^2 |\mathbf{q}|^2}.
\end{align}

For a long flat-top pulse (plateau) with $E(t)=E_0$ and $A(t)=-E_0 t$ over the relevant time window, the QKE system reproduces the standard Landau--Zener (Schwinger-like) result for the final conduction-band occupation (see, e.g., Refs.~\cite{dora_prb_2010,gavrilov_prd_2012,akal_prd_2016} and references therein),
\begin{equation}
  n_{\mathbf{p}} \simeq \exp\!\left[-\pi\,\frac{\Delta^2+(\hbar v_\text{F} k_\perp)^2}{\hbar v_\text{F} |e|E_0}\right],
\end{equation}
providing a nontrivial check of units and sign conventions. In the quasistatic regime $\gamma_\text{D}=\Delta/(v_\text{F} |e| E_0\tau)\ll 1$, the peak values of the numerically computed spectra approach this exponential estimate.

\vspace{0.2cm}

\begin{center}
{\bf D. ENERGY-RESOLVED CARRIER SPECTRUM: REQUIREMENTS FOR INTERBAND PROBES}
\end{center}

\vspace{0.1cm}

A standard far-field readout of nonequilibrium carriers in graphene-based systems is pump--probe spectroscopy in the THz-to-mid-IR range, where the pump-induced change of absorption/reflectivity (or equivalently of the complex optical conductivity) is measured as a function of probe photon energy $\hbar\Omega$~\cite{dawlaty_apl_2008,horng_prb_2011,jnawali_nano_lett_2013}. Interband probes sample momentum-shell averages set primarily by the resonance condition $2\varepsilon_{\mathbf{k}}\simeq\hbar\Omega$ and are therefore naturally sensitive to structured post-pulse populations. In this section we recast our post-pulse momentum distribution $n_{\mathbf{k}}$ into an energy-resolved spectrum and extract the characteristic energy scale of the oscillations, which directly determines the spectral resolution needed to resolve the corresponding modulation in energy-resolved pump--probe observables.

We consider the gapped Dirac dispersion
\begin{equation}
\varepsilon_{\mathbf{k}}=\sqrt{\Delta^2+(\hbar v_\text{F} |\mathbf{k}|)^2},
\label{eq:SM_epsk}
\end{equation}
and denote by $n_{\mathbf{k}}$ the post-pulse occupation of the conduction band (electron--hole symmetric excitation implies an identical hole population in the valence band). The total number of excited carriers in an illuminated area $S = L^2$ is
\begin{equation}
N = g S \int \! \frac{d^2\mathbf{k}}{(2\pi)^2}\,n_{\mathbf{k}},
\label{eq:SM_N_def}
\end{equation}
where $g=4$ accounts for spin and valley degeneracy. Introducing polar coordinates and the angular average
\begin{equation}
\overline{n}(|\mathbf{k}|)\equiv \frac{1}{2\pi}\int\limits_{0}^{2\pi} \! n_{\mathbf{k}} \, d\varphi_\mathbf{k},
\label{eq:SM_nbar_def}
\end{equation}
we define the energy-resolved spectrum by
\begin{equation}
\frac{dN}{d\varepsilon} = \frac{S\Delta}{(\hbar v_\text{F})^2} \, \mathcal{N} (\varepsilon),
\label{eq:dN-deps}
\end{equation}
where $\mathcal{N} (\varepsilon)$ is a dimensionless energy-resolved distribution per unit area up to the prefactor,
\begin{equation}
\mathcal{N} (\varepsilon) = \frac{g}{2\pi}\,\frac{\varepsilon}{\Delta}\,
\overline{n} \big(|\mathbf{k}|(\varepsilon)\big),\qquad |\mathbf{k}|(\varepsilon) = \frac{\sqrt{\varepsilon^2 - \Delta^2}}{\hbar v_\text{F}},
\label{eq:energy_N_eps}
\end{equation}
and $\varepsilon \geqslant \Delta$. The dimensionless function $\mathcal{N} (\varepsilon)$ provides a direct mapping from our computed $n_{\mathbf{k}}$ to an energy-resolved distribution.

\begin{figure}[h]
\center{\includegraphics[width=0.5\linewidth]{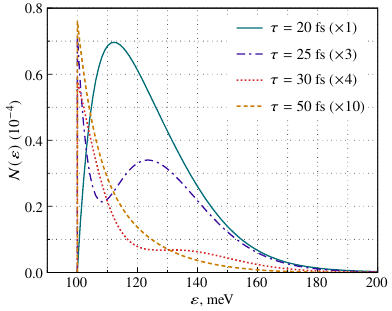}}
\caption{Energy-resolved excited-carrier spectrum $\mathcal{N} (\varepsilon)$ [Eq.~\eqref{eq:energy_N_eps}] for the deformed single-lobe pulse $e(z)=\mathrm{exp} (-z^{2}-z^{4})$ in gapped Dirac graphene at fixed $E_0=40~\text{kV}/\text{cm}$ and $\Delta=0.1~\text{eV}$. Curves correspond to pulse durations $\tau=20$, $25$, $30$, and $50~\text{fs}$, i.e. $\gamma_{\text{D}}=1.25$, $1.0$, $0.833$, and $0.5$, respectively. As $\gamma_{\text{D}}$ increases through the nonadiabatic crossover, the energy distribution develops pronounced oscillations with characteristic spacing $\delta\varepsilon_\text{fr} \sim 20~\text{meV}$, providing the characteristic spectral scale needed to detect the associated modulation in energy-resolved pump--probe observables.}
\label{fig:SM_energyDist}
\end{figure}

To assess experimental feasibility, we analyze the oscillatory structure of $\mathcal{N} (\varepsilon)$ for the Gaussian and deformed single-lobe pulses used in the main text and extract the characteristic energy spacing $\delta\varepsilon_\text{fr}$ between neighboring extrema in $\varepsilon$. Since interband probes are governed by the resonance condition $2\varepsilon\simeq\hbar\Omega$, oscillations on the scale $\delta\varepsilon_\text{fr}$ translate into modulation of pump-induced absorption/reflectivity change on the photon-energy scale $\delta(\hbar\Omega)\simeq 2 \delta\varepsilon_\text{fr}$. Therefore, resolving the corresponding modulation in the energy-resolved pump--probe signal requires an effective spectral resolution
\begin{equation}
\Delta(\hbar\Omega)\ \lesssim\ 2 \delta\varepsilon_\text{fr}.
\label{eq:SM_resolution}
\end{equation}
In Fig.~\ref{fig:SM_energyDist} we plot $\mathcal{N} (\varepsilon)$ for representative parameters in the nonadiabatic crossover regime and find $\delta\varepsilon_\text{fr}\sim 20~\text{meV}$, implying a resolution requirement $\Delta(\hbar\Omega)\lesssim 40~\text{meV}$. Such resolution is compatible with narrowband ultrafast probes and spectrally resolved transient absorption/reflectivity measurements in the THz-to-mid-IR range~\cite{dawlaty_apl_2008,horng_prb_2011,jnawali_nano_lett_2013}. We also note that the small values $\mathcal{N}(\varepsilon)\sim 10^{-5}$ in Fig.~\ref{fig:SM_energyDist} correspond to a non-negligible absolute yield once the prefactor in Eq.~\eqref{eq:dN-deps} is restored. For an illuminated spot of linear size $L\sim 10$--$100~\qty{}{\micro\metre}$ (area $S=L^2$) and $\Delta=0.1~\text{eV}$, one finds
\begin{equation*}
\frac{S \Delta}{(\hbar v_\text{F})^2} \sim (10^7 \text{--} 10^9)~\text{eV}^{-1},
\end{equation*}
so that $(1/S)\,dN/d\varepsilon$ can be sizable even when $\mathcal{N}(\varepsilon)$ is numerically small.

Finally, it is useful to relate the required spectral resolution to the characteristic duration of a transform-limited probe pulse. For a Gaussian probe one has the time--bandwidth product $\Delta\nu_{\rm FWHM}\Delta t_{\rm FWHM}\simeq 0.44$, with $\Delta E_{\rm FWHM}=h\,\Delta\nu_{\rm FWHM}$. Thus a spectral resolution $\Delta(\hbar\Omega)\sim 40~\text{meV}$ corresponds to a pulse duration $\Delta t_{\rm FWHM}\approx 0.44\,h/\Delta E \simeq 4.5\times 10^{-14}~\text{s}\approx 45~\text{fs}$. This is many orders of magnitude shorter than the ballistic escape time across a far-field spot, $t_\text{esc}\sim L/v_\text{F}\sim 10$--$100~\text{ps}$ for $L\sim 10$--$100~\qty{}{\micro\metre}$, so transport out of the driven region is negligible on the pump--probe timescales relevant for resolving the predicted modulation.

\newpage

\begin{center}
{\bf E. SPECTRAL FRINGES FOR ZERO-AREA PULSES}
\end{center}

\vspace{0.1cm}

A strictly single-lobe electric field with a nonvanishing time integral is not generic for freely propagating far-field transients, since Maxwell constraints and source dynamics typically imply that any dominant lobe is accompanied by weaker components that restore (approximately or exactly) zero net field area when integrated over sufficiently long times. In practice, waveform-asymmetric single-/half-cycle transients commonly consist of a dominant short lobe followed by a weaker, longer compensating tail of opposite sign; such quasi-half-cycle waveforms are standard in photoconductive-antenna sources and related THz emitters~\cite{schmuttenmaer_2004,isgandarov_2021}. To demonstrate that our mechanism does not rely on a nonzero field area, we enforce an explicitly zero-area pulse by the substitution
\begin{equation}
e(z)\ \to\ e_{\eta}(z)=e(z)-\eta e(\eta z),
\label{eq:e_eta}
\end{equation}
where $\eta$ controls the duration $T=\tau/\eta$ of the compensating tail. This construction preserves the dominant lobe while guaranteeing
\begin{equation}
\int\limits_{-\infty}^{\infty} \! E_{\eta}(t) dt = 0 \qquad (\eta>0).
\end{equation}

\begin{figure}[h]
\centering
\includegraphics[width=0.6\linewidth]{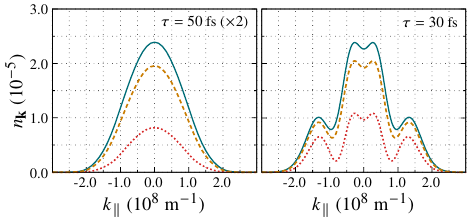}
\caption{Longitudinal momentum distributions of the carriers in gapped graphene for $k_\perp=0$, $\Delta=0.1~\text{eV}$ and $E_0=40~\text{kV}/\text{cm}$, driven by explicitly zero-area pulses~\eqref{eq:e_eta} built from the deformed envelope
$e(z) = \mathrm{exp}(-z^2-z^4)$. Curves correspond to $\eta=0$ (solid), $\eta=0.02$ (dashed), and $\eta=0.1$ (dotted). The main-lobe duration is (left)~$\tau=50~\text{fs}$ and (right)~$\tau=30~\text{fs}$. For visibility, the data in panel~(a) are multiplied by a factor of $2$.}
\label{fig:SM_zero-area}
\end{figure}

Figure~\ref{fig:SM_zero-area} compares longitudinal momentum spectra in gapped graphene for several values of $\eta$. Panel~(a) confirms that in the smooth regime at relatively large $\tau$ (small $\gamma_{\text{D}}$) the spectra remain single-peaked even after enforcing zero net area of the external field. Panel~(b) shows that the pronounced fringe pattern for a shorter main lobe persists for all tested zero-area pulses and becomes essentially $\eta$-independent for $\eta\lesssim 0.02$ (i.e. $T\gtrsim 50 \tau$), where the spectra nearly coincide pointwise with the $\eta=0$ results of the main text. The range $\eta\sim 0.02$--$0.1$ used here is representative of experimentally generated quasi-half-cycle waveforms, in which a dominant lobe is followed by a longer, weaker compensating tail~\cite{schmuttenmaer_2004,isgandarov_2021}.

\begin{thebibliography}{99}
%
\bibitem{sauter_1931}
F.~Sauter,
{\"Uber das Verhalten eines Elektrons im homogenen elektrischen Feld nach der relativistischen Theorie Diracs},
{Z.~Phys.} {\bf 69}, 742 (1931).
%
\bibitem{heisenberg_euler}
W.~Heisenberg and H.~Euler,
{Folgerungen aus der Diracschen Theorie des Positrons},
{Z.~Phys.} {\bf 98}, 714 (1936).
%
\bibitem{weisskopf}
V.~Weisskopf,
{\"Uber die Elektrodynamik des Vakuums auf Grund der Quantentheorie des Elektrons},
{Kong. Dan. Vid. Sel. Mat. Fys. Med.} {\bf 14N6}, 1 (1936).
%
\bibitem{schwinger_1951}
J.~Schwinger,
{On gauge invariance and vacuum polarization},
{Phys. Rev.} {\bf 82}, 664 (1951).
%
\bibitem{dipiazza_rmp_2012}
A.~Di Piazza, C.~M\"uller, K.~Z.~Hatsagortsyan, and C.~H.~Keitel,
{Extremely high-intensity laser interactions with fundamental quantum systems},
{Rev. Mod. Phys.} {\bf 84}, 1177 (2012).
%
\bibitem{xie_review_2017}
B.~S.~Xie, Z.~L.~Li, and S.~Tang,
{Electron-positron pair production in ultrastrong laser fields},
{Matter Radiat. Extremes} {\bf 2}, 225 (2017).
%
\bibitem{gonoskov_2022} A.~Gonoskov, T.~G.~Blackburn, M.~Marklund, and S.~S.~Bulanov, Charged particle motion and radiation in strong electromagnetic fields, Rev. Mod. Phys. {\bf 94}, 045001 (2022).
%
\bibitem{fedotov_review}
A.~Fedotov, A.~Ilderton, F.~Karbstein, B.~King, D.~Seipt, H.~Taya, and G.~Torgrimsson,
{Advances in QED with intense background fields},
{Phys. Rep.} {\bf 1010}, 1 (2023).
%
\bibitem{popruzhenko_ufn_2023} S.~V.~Popruzhenko and A.~M.~Fedotov, Dynamics and
radiation of charged particles in ultra-intense laser fields, Phys. Usp. {\bf 66}, 460 (2023).
%
\bibitem{hebenstreit_prl_2009} F.~Hebenstreit, R.~Alkofer, G.~V.~Dunne, and H. Gies, Momentum signatures for Schwinger pair production in short laser pulses with a subcycle structure, Phys. Rev. Lett. {\bf 102}, 150404 (2009).
%
\bibitem{dumlu_dunne_prl_2010} C.~K.~Dumlu and G.~V.~Dunne, Stokes phenomenon and Schwinger vacuum pair production in time-dependent laser pulses, Phys. Rev. Lett. {\bf 104}, 250402 (2010).
%
\bibitem{dumlu_dunne_prd_2011_1} C.~K.~Dumlu and G.~V.~Dunne, Interference effects in Schwinger vacuum pair production for time-dependent laser pulses, Phys. Rev. D {\bf 83}, 065028 (2011).
%
\bibitem{dumlu_dunne_prd_2011_2} C.~K.~Dumlu and G.~V.~Dunne, Complex worldline instantons and quantum interference in vacuum pair production, Phys. Rev. D {\bf 84}, 125023 (2011).
%
\bibitem{akkermans_prl_2012} E.~Akkermans and G.~V.~Dunne, Ramsey fringes and time-domain multiple-slit interference from vacuum, Phys. Rev. Lett. {\bf 108}, 030401 (2012).
%
\bibitem{abdukerim_plb_2013} N.~Abdukerim, Z.~Li, and B.~S.~Xie, Effects of laser pulse shape and carrier envelope phase on pair production, Phys. Lett. B {\bf 726}, 820 (2013).
%
\bibitem{kohlfuerst_prd_2013} C.~Kohlf\"urst, M.~Mitter, G.~von Winckel, F.~Hebenstreit, and R.~Alkofer, Optimizing the pulse shape for Schwinger pair production, Phys. Rev. D {\bf 88}, 045028 (2013).
%
\bibitem{hebenstreit_plb_2014} F.~Hebenstreit and F.~Fillion-Gourdeau, Optimization of Schwinger pair production in colliding laser pulses, Phys. Lett. B {\bf 739}, 189 (2014).
%
\bibitem{linder_prd_2015} M.~F.~Linder, C.~Schneider, J.~Sicking, N.~Szpak, and R.~Sch\"utzhold, Pulse shape dependence in the dynamically assisted Sauter-Schwinger effect, Phys. Rev. D {\bf 92}, 085009 (2015).
%
\bibitem{aleksandrov_prd_2017} I.~A.~Aleksandrov, G.~Plunien, and V.~M.~Shabaev, Pulse shape effects on the electron-positron pair production in strong laser fields, Phys. Rev. D {\bf 95}, 056013 (2017).
%
\bibitem{allor_prd_2008} D.~Allor, T.~D.~Cohen, and D.~A.~McGady, The Schwinger mechanism and graphene, Phys. Rev. D \textbf{78}, 096009 (2008).
%
\bibitem{dora_prb_2010} B.~D\'ora and R.~Moessner, Nonlinear electric transport in graphene: Quantum quench dynamics and the Schwinger mechanism, Phys. Rev. B {\bf 81}, 165431 (2010).
%
\bibitem{lewkowicz_prb_2011} M.~Lewkowicz, H.~C.~Kao, and B.~Rosenstein, Signature of Schwinger’s pair creation rate via radiation generated in graphene by strong electric current, Phys. Rev. B {\bf 84}, 035414 (2011).
%
\bibitem{gavrilov_prd_2012} S.~P.~Gavrilov, D.~M.~Gitman, and N.~Yokomizo, Dirac fermions in strong electric field and quantum transport in graphene, Phys. Rev. D {\bf 86}, 125022 (2012).
%
\bibitem{akal_prd_2016} I.~Akal, R.~Egger, C.~M\"uller, and S.~Villalba-Ch\'avez, Low-dimensional approach to pair production in an oscillating electric field: Application to bandgap graphene layers, Phys. Rev. D \textbf{93}, 116006 (2016).
%
\bibitem{linder_prb_2018} M.~F.~Linder, A.~Lorke, and R.~Sch\"utzhold, Analog Sauter--Schwinger effect in semiconductors for spacetime-dependent fields, Phys. Rev. B \textbf{97}, 035203 (2018).
%
\bibitem{akal_prd_2019} I.~Akal, R.~Egger, C.~M\"uller, and S.~Villalba-Ch\'avez, Simulating dynamically assisted production of Dirac pairs in gapped graphene monolayers, Phys. Rev. D \textbf{99}, 016025 (2019).
%
\bibitem{zhou_nat-mat_2007} S.~Zhou {\it et al.}, Substrate-induced bandgap opening in epitaxial graphene. Nature Mater {\bf 6}, 770 (2007). 
%
\bibitem{kim_prl_2008} S.~Kim, J.~Ihm, H.~J.~Choi, and Y.-W.~Son, Origin of anomalous electronic structures of epitaxial graphene on silicon carbide, Phys. Rev. Lett. {\bf 100}, 176802 (2008).
%
\bibitem{enderlein_njp_2010} C.~Enderlein, Y.~S.~Kim, A.~Bostwick, E.~Rotenberg, and K.~Horn, The formation of an energy gap in graphene on ruthenium by controlling the interface, New J. Phys. {\bf 12}, 033014 (2010).
%
\bibitem{yu_apl_2013} C.~Yu {\it et al.}, Buffer layer induced band gap and surface low energy optical phonon scattering in epitaxial graphene on SiC(0001), Appl. Phys. Lett. {\bf 102}, 013107 (2013).
%
\bibitem{riedl_prl_2009} C.~Riedl, C.~Coletti, T.~Iwasaki, A.~A.~Zakharov, and U.~Starke, Quasi-free-standing epitaxial graphene on SiC obtained by hydrogen intercalation, Phys. Rev. Lett. {\bf 103}, 246804 (2009).
%
\bibitem{bianco_apl_2015} F.~Bianco, D.~Perenzoni, D.~Convertino, S.~L.~De Bonis, D.~Spirito, M.~Perenzoni, C.~Coletti, M.~S.~Vitiello, and A.~Tredicucci, Terahertz detection by epitaxial-graphene field-effect-transistors on silicon carbide, Appl. Phys. Lett. {\bf 107}, 131104 (2015).
%
\bibitem{santos_sci-rep_2016} C.~N.~Santos, F.~Joucken, D.~De Sousa Meneses, P.~Echegut, J.~Campos-Delgado, P.~Louette, J.-P,~Raskin, and B.~Hackens, Terahertz and mid-infrared reflectance of epitaxial graphene, Sci. Rep. {\bf 6}, 24301 (2016).
%
\bibitem{paschke_aqt_2020}  F.~Paschke, T.~Birk, S.~Forti, U.~Starke, and M.~Fonin, Hydrogen-intercalated graphene on SiC as platform for hybrid superconductor devices, Adv. Quantum Technol. {\bf 3}, 2000082 (2020).
%
\bibitem{singh_jpd_2024} A.~Singh, H.~N\v{e}mec, J.~Kunc, and P.~Ku\v{z}el, Ultrafast terahertz conductivity in epitaxial graphene nanoribbons: an interplay between photoexcited and secondary hot carriers, J. Phys. D: Appl. Phys. {\bf 58}, 045307 (2024).
%
\bibitem{hafez_opt-mat_2020} H.~A.~Hafez, S.~Kovalev, K.-J.~Tielrooij, M.~Bonn, M.~Gensch, and D.~Turchinovich, Terahertz nonlinear optics of graphene: From saturable absorption to high-harmonics generation, Adv. Opt. Mater. {\bf 8}, 1900771 (2020).
%
\bibitem{oladyshkin_prb_2017} I.~V.~Oladyshkin, S.~B.~Bodrov, Yu.~A.~Sergeev, A.~I.~Korytin, M.~D.~Tokman, and A.~N.~Stepanov, Optical emission of graphene and electron-hole pair production induced by a strong terahertz field, Phys. Rev. B {\bf 96}, 155401 (2017).
%
\bibitem{wallbank_prb_2013_1} J.~R.~Wallbank, A.~A.~Patel, M.~Mucha-Kruczy\ifmmode \acute{n}\else \'{n}\fi{}ski, A.~K.~Geim, and V.~I.~Fal'ko, Generic miniband structure of graphene on a hexagonal substrate, Phys. Rev. B {\bf 87}, 245408 (2013).
%
\bibitem{wallbank_prb_2013_2} J.~R.~Wallbank, M.~Mucha-Kruczy\ifmmode \acute{n}\else \'{n}\fi{}ski, and V.~I.~Fal'ko, Moir\'e minibands in graphene heterostructures with almost commensurate $\sqrt{3}\ifmmode\times\else\texttimes\fi{}\sqrt{3}$ hexagonal crystals, Phys. Rev. B {\bf 88}, 155415 (2013).
%
\bibitem{wallbank_adp_2015} J.~R.~Wallbank, M.~Mucha-Kruczy\ifmmode \acute{n}\else \'{n}\fi{}ski, X.~Chen, and V.~I.~Fal'ko, Moir\'e superlattice effects in graphene/boron‐nitride van der Waals heterostructures, Ann. Phys. {\bf 527}, 359 (2015).
%
\bibitem{jung_prb_2017} J.~Jung, E.~Laksono, A.~M.~DaSilva, A.~H.~MacDonald, M.~Mucha-Kruczy\ifmmode \acute{n}\else \'{n}\fi{}ski, and S.~Adam, Moir\'e band model and band gaps of graphene on hexagonal boron nitride, Phys Rev. B {\bf 96}, 085442 (2017).
%
\bibitem{li_comm-phys_2020} Y.~Li, M.~Amado, T.~Hyart, G.~P.~Mazur, and J.~W.~A.~Robinson, Topological valley currents via ballistic edge modes in graphene superlattices near the primary Dirac point, Commun. Phys. {\bf 3}, 224 (2020).
%
\bibitem{aleksandrov_prd_2016} I.~A.~Aleksandrov, G.~Plunien, and V.~M.~Shabaev, Electron-positron pair production in external electric fields varying both in space and time, Phys. Rev. D {\bf 94}, 065024 (2016).
%
\bibitem{popov_2005} V.~S.~Popov, Imaginary-time method in quantum mechanics and field theory, Phys. At. Nucl. {\bf 68}, 686 (2005).
%
\bibitem{oertel_prd_2019} J.~Oertel and R.~Sch\"utszhold, WKB approach to pair creation in spacetime-dependent fields: The case of a spacetime-dependent mass, Phys. Rev. D {\bf 99}, 125014 (2019).
%
\bibitem{taya_jhep_2021} H.~Taya, T.~Fujimori, T.~Misumi, M.~Nitta, and N.~Sakai, Exact WKB analysis of the vacuum pair production by time-dependent electric fields, J. High Energy Phys. 03 (2021) 082.
%
\bibitem{SM} See Supplemental Material at [URL to be inserted] for additional theoretical details, numerical checks, and supporting figures.
%
\bibitem{smolyansky_particles_2019} S.~A.~Smolyansky, A.~D.~Panferov, D.~B.~Blaschke, and N.~T.~Gevorgyan, Nonperturbative kinetic description of electron-hole excitations in graphene in a time dependent electric field of arbitrary polarization, Particles {\bf 2}, 208 (2019).
%
\bibitem{smolyansky_particles_2020} S.~A.~Smolyansky, A.~D.~Panferov, D.~B.~Blaschke, and N.~T.~Gevorgyan, Kinetic equation approach to graphene in strong external fields, Particles {\bf 3}, 456 (2020).
%
\bibitem{dawlaty_apl_2008} J.~M.~Dawlaty, S.~Shivaraman, M.~Chandrashekhar, F.~Rana, and M.~G.~Spencer, Measurement of the optical absorption spectra of epitaxial graphene from terahertz to visible, Appl. Phys. Lett. {\bf 93}, 131905 (2008).
%
\bibitem{horng_prb_2011} J.~Horng, C.-F.~Chen, B.~Geng, C.~Girit, Y.~Zhang, Z.~Hao, H.~A.~Bechtel, M.~Martin, A.~Zettl, M.~F.~Crommie, Y.~R.~Shen, and F.~Wang, Drude conductivity of Dirac fermions in graphene, Phys. Rev. B {\bf 83}, 165113 (2011).
%
\bibitem{jnawali_nano_lett_2013} G.~Jnawali, Y.~Rao, H.~Yan, and T.~F.~Heinz, Observation of a transient decrease in terahertz conductivity of photoexcited graphene, Nano Lett. {\bf 13}, 524 (2013).
%
\bibitem{gavrilov_universe_2020} S.~P.~Gavrilov, D.~M.~Gitman, V.~V.~Dmitriev, A.~D.~Panferov, and S.~A.~Smolyansky, Radiation problems accompanying carrier production by an electric field in the graphene, Universe {\bf 6}, 205 (2020).
%
\bibitem{schmuttenmaer_2004} C.~A.~Schmuttenmaer, Exploring dynamics in the far-infrared with terahertz spectroscopy, Chem. Rev. {\bf 104}, 1759 (2004).
%
\bibitem{isgandarov_2021} E.~Isgandarov, X.~Ropagnol, M.~Singh, and T.~Ozaki, Intense terahertz generation from photoconductive antennas, Front. Optoelectron. {\bf 14}, 64 (2021).
%
\end{thebibliography}
\end{document}